\documentclass[prd,aps,preprintnumbers,nofootinbib,superscriptaddress]{revtex4}
\usepackage{amssymb,amsmath,graphicx,bm}

\allowdisplaybreaks
\usepackage[usenames,dvipsnames]{color}
\usepackage[normalem]{ulem} 
\usepackage{slashed} 
\usepackage[caption=false]{subfig} 
\renewcommand\sout{\bgroup \color[rgb]{0.55,0.00,0.99} \ULdepth=-.5ex \ULset}
\newcommand{\xB }{x_{\scriptscriptstyle B}}
\newcommand{\sT}{{\scriptscriptstyle T}}
\renewcommand{\d}{\mathrm{d}}
\def\slash#1{\setbox0=\hbox{$#1$}               
        \dimen0=\wd0                            
        \setbox1=\hbox{/} \dimen1=\wd1          
        \ifdim\dimen0>\dimen1                   
        \rlap{\hbox to \dimen0{\hfil/\hfil}}    
        #1                                      
        \else
        \rlap{\hbox to \dimen1{\hfil$#1$\hfil}} 
        /                                       
        \fi}                                    %
\begin{document}

\title{$J/\psi$ meson production in SIDIS: matching  high and low transverse momentum}

\author{Dani\"el Boer}
\email{d.boer@rug.nl}
\affiliation{ {Van Swinderen Institute for Particle Physics and Gravity, University of Groningen, Nijenborgh 4, 9747 AG Groningen, The Netherlands}}

\author{Umberto D'Alesio}
\email{umberto.dalesio@ca.infn.it}
\affiliation{Dipartimento di Fisica, Universit\`a di Cagliari, Cittadella Universitaria, I-09042 Monserrato, Cagliari, Italy}
\affiliation{INFN, Sezione di Cagliari, Cittadella Universitaria, I-09042 Monserrato, Cagliari, Italy}

\author{Francesco Murgia}
\email{francesco.murgia@ca.infn.it}
\affiliation{INFN, Sezione di Cagliari, Cittadella Universitaria, I-09042 Monserrato, Cagliari, Italy}

\author{Cristian Pisano}
\email{cristian.pisano@unica.it}
\affiliation{Dipartimento di Fisica, Universit\`a di Cagliari, Cittadella Universitaria, I-09042 Monserrato, Cagliari, Italy}
\affiliation{INFN, Sezione di Cagliari, Cittadella Universitaria, I-09042 Monserrato, Cagliari, Italy}

\author{Pieter Taels}
\email{pieter.taels@polytechnique.edu}
\affiliation{INFN, Sezione di Cagliari, Cittadella Universitaria, I-09042 Monserrato, Cagliari, Italy}
\affiliation{Centre de Physique Th\'eorique, \'Ecole polytechnique, CNRS, I.P.\ Paris, F-91128 Palaiseau, France}

\begin{abstract}
\end{abstract}
\date{\today}

\begin{abstract}
We consider the transverse momentum spectrum and the $\cos 2\phi$ azimuthal distribution of $J/\psi$ mesons produced in semi-inclusive, deep-inelastic electron-proton scattering, where the electron and the proton are unpolarized. At low transverse momentum, we propose factorized expressions in terms of transverse momentum dependent gluon distributions and shape functions. We show that our formulae, at the order $\alpha_s,$ correctly match with the collinear factorization results at high transverse momentum. The latter are computed at the order $\alpha_s^2$ in the framework of nonrelativistic QCD (NRQCD),  with the inclusion of the intermediate $^3S_1^{[1]}$ color-singlet Fock state, as well as the subleading  color-octet ones that are relatively suppressed by a factor $v^4$ in the NRQCD velocity parameter $v$. We show that the  $^1\!S_0^{[8]}$ and $^3\!P_J^{[8]}$ ($J = 0,1,2$) contributions diverge in the small transverse momentum region and allow us to determine the perturbative tails of the shape functions, which carry the same quantum numbers. These  turn out to be identical, except for the overall magnitude given by the appropriate NRQCD long distance matrix element.
\end{abstract}

\maketitle

\section{Introduction}

The production of a light hadron $h$ with a specific transverse momentum in semi-inclusive, deep-inelastic electron-proton scattering (SIDIS), $e\,p\to e^\prime\, h\,X$, is in general characterized by three different scales: the hard scale of the process $Q$, given by the virtuality of the gauge boson exchanged in the reaction,  the nonperturbative QCD scale $\Lambda_{\rm QCD}$, and the magnitude of the hadron transverse momentum $q_\sT$ in a suitable reference frame. Depending on the value of $q_\sT$, two different factorization frameworks can be  adopted for the description of this process. Both of them enable  to separate the short-distance from the long-distance contributions to the cross section. While the former can be perturbatively calculated through a systematic expansion in the strong coupling constant, the latter has to be parametrized in terms of parton distributions (PDFs) and fragmentation functions (FFs), which need to be extracted from data. 

More explicitly, collinear factorization is applicable in the so-called high-$q_T$ region, namely  for $q_T \gg \Lambda_{\rm QCD}$,
where the transverse momentum in the final state is generated by  perturbative radiation and the cross section is expressed in terms
of collinear ({\it i.e.}, integrated over transverse momentum) PDFs and FFs. The other framework, based on transverse momentum dependent (TMD) factorization~\cite{Collins:2011zzd,GarciaEchevarria:2011rb,Echevarria:2012js} is valid  at low $q_T$, $q_T \ll Q$,  and involves TMD PDFs and FFs (or TMDs for short). The high- and low-$q_\sT$ regions overlap for $\Lambda_{\rm QCD} \ll  q_\sT \ll Q$, where  both descriptions  can therefore be applied. If the two results describe the same dynamics, characterized  by the same power behavior, they have to match in this intermediate region. Conversely, if  the two results describe competing mechanisms, they should be considered independently and added together~\cite{Bacchetta:2008xw}. 

The SIDIS cross section differential in $q_\sT$ and integrated over the azimuthal angle of the final hadron can be expressed in terms of  unpolarized, twist-two TMDs in the small-$q_\sT$ region, and its matching with the collinear description has been demonstrated in Ref.~\cite{Bacchetta:2008xw}.  The matching for the analogous observable in Drell-Yan (DY) dilepton production,  $p\, p\to \ell\, \ell^\prime\,X$, has been proven as well in Refs.~\cite{Collins:1984kg,Catani:2000vq}.  
Azimuthal asymmetries in both SIDIS~\cite{Bacchetta:2006tn} and DY~\cite{Boer:2006eq,Berger:2007jw,Chen:2016hgw} processes have also been widely investigated within the TMD framework. In particular, the matching of the $\cos\phi$ modulations, which  are suppressed by a factor $q_\sT/Q$ with respect to the $\phi$-integrated cross sections and involve twist-three TMDs, has been shown very recently  in Ref.~\cite{Bacchetta:2019qkv}.

In this paper, we analyze $J/\psi$ production in SIDIS, $e\, p \to e^\prime\, J/\psi\, X$,  along the same lines of Refs.~\cite{Bacchetta:2008xw,Bacchetta:2019qkv}. There are several reasons that motivate our study. First of all,  since its discovery in 1974, the $J/\psi$ meson, a charm-anticharm quarkonium bound state with odd charge parity, has always attracted a lot of attention as a probe of the perturbative and nonperturbative aspects of quantum chromodynamics (QCD) and their interplay. Moreover, $J/\psi$ production in both $ep$~\cite{Yuan:2008vn,Bacchetta:2018ivt,Mukherjee:2016qxa,DAlesio:2019qpk} and $pp$ collisions~\cite{Yuan:2008vn,Dunnen:2014eta,Lansberg:2017tlc,Lansberg:2017dzg,Scarpa:2019fol} has been proposed lately as a tool to access gluon TMDs. Similar studies have also been carried out within the so-called generalized parton model approach~\cite{Godbole:2013bca,Godbole:2014tha,Godbole:2017syo,Kishore:2018ugo,Rajesh:2018qks,DAlesio:2017rzj,DAlesio:2018rnv,DAlesio:2019gnu}. From the experimental point of view, these reactions should have a very clean signature due to the large branching ratio of the $J/\psi$  leptonic decay modes. Hence our findings could be in principle verified at the future Electron-Ion Collider (EIC) planned in the United States~\cite{Boer:2011fh,Accardi:2012qut}.

As compared to $e\,p\to e^\prime\, h\,X$,  the study of  $e\, p \to e^\prime\, J/\psi\, X$ presents an additional complication, namely  a second hard scale given by the  $J/\psi$ mass, $M_\psi$. Since we want to avoid contributions from photoproduction processes, we focus on the kinematic region where $Q \gtrsim M_\psi$.  In principle, each of the two scales, or any combination of them, can be chosen as the factorization scale $\mu$ in the calculation of the cross section. From our analysis it will turn out  that  the choice $\mu = \sqrt{Q^2+M_\psi^2}$ allows for a smooth transition of the cross section from the high- to low-$q_\sT$ region.  

When the $J/\psi$ meson is produced with a large transverse momentum,  $q_\sT \gg \Lambda_{\rm QCD}$, a collinear factorization approach based on fixed-order perturbative QCD can be applied. Moreover, for the description of the production mechanism of the quarkonium state we rely on nonrelativistic QCD (NRQCD). This rigorous theoretical framework implies a separation of short-distance coefficients, which  can  be  calculated  perturbatively  as expansions in the strong-coupling constant $\alpha_s$, from long-distance matrix elements (LDMEs)  $\langle 0 \vert {\cal O}(n)\vert 0\rangle$,  which must be extracted from experiment~\cite{Bodwin:1994jh}.  In the definition of the LDMEs,  $n= \,^{\! 2S+1} L_J ^{[c]}$, where $S$ denotes the spin of the produced charm-anticharm quark pair, $L$  the 
orbital angular momentum, $J$ the total angular momentum and $c$ the color configuration, with $c = 1, 8$. The relative importance of the LDMEs can be estimated by  means of  velocity  scaling  rules,  \textit{i.e.} they  are predicted  to  scale  with a definite power of the heavy quark-antiquark  relative velocity $v$ in the limit $ v\ll 1$. For charmonium states $v^2 \sim 0.3$, whereas for bottomonium ones $v^2 \sim 0.1$. In this way, the theoretical predictions are organized as double expansions in $\alpha_s$  and $v$. The main feature of this formalism is that the charm-anticharm quark pair forming the bound state can be produced both in a color-singlet (CS) configuration, with the same quantum numbers as the $J/\psi$ meson, and as an intermediate color-octet (CO) state with different quantum numbers. In the latter case, the pair subsequently evolves into a  colorless state through the emission of soft gluons. For an $S$-wave quarkonium state like the $J/\psi$ and $\Upsilon$ mesons, the main contribution in the $v$ expansion is given by the $^3S_1^{[1]}$ CS state, and in the limit $v \to 0$ NRQCD reduces to the traditional color-singlet model.  The $^1S_0^{[8]}$, $^3S_1^{[8]}$, $^3P_J^{[8]}$ (with $J=0,1,2$) CO states contribute to the leading relativistic corrections, and the corresponding LDMEs are suppressed by a factor $v^4$ relative to the CS matrix element. While the latter can be determined from potential models, lattice calculations or from leptonic decays, the CO ones are usually extracted from fits to data on $J/\psi$ and $\Upsilon$ yields and, at present, the knowledge of them is not very accurate. Furthermore, although NRQCD successfully explains many experimental observations, like inclusive charmonium production in $\bar p p $ collisions at the Tevatron, it has problems in reproducing all charmonium cross sections and polarization measurements from different processes  in a consistent way. For a recent review on the subject, see Ref.~\cite{Lansberg:2019adr}. 
f
In the small-$q_\sT$ region, $q_\sT\ll M_\psi$,  TMD factorization has not yet been  proven in a rigorous way for  the process $e\,p \to e^\prime \,J/\psi\,X$. There are  however strong arguments in favor of its validity, if we consider the analogy with $e\,p \to e^\prime \, h \,X$, for which TMD factorization holds at all orders~\cite{Collins:2011zzd}. The only difference from the color point of view is that the dominant partonic subprocess is now $\gamma^* g \to c\bar c$ instead of $\gamma^* q \to q^\prime$. Hence final state interactions will be resummed in the gauge link of the gluon correlator, which will be in the adjoint representation, rather than in a quark correlator in the fundamental representation. Since the $J/\psi$ mass does not affect the gauge link structure, we do not expect any TMD factorization breaking effects due to color entanglement~\cite{Bacchetta:2018ivt}. 

Motivated by these arguments, in the present analysis we put forward factorization formulae, valid at the twist-two level, for the transverse momentum spectrum and the  $\cos2\phi$ azimuthal distribution of $J/\psi$ mesons produced in SIDIS.  In addition to the usual TMD PDFs, we consider the recently proposed shape functions~\cite{Echevarria:2019ynx,Fleming:2019pzj}, which are the generalization of the collinear LDMEs in NRQCD. Alternatively, they can be seen as the analog of the TMD FFs for light hadron production in SIDIS.  By requiring a proper matching with the collinear results, in complete analogy with the TMD cross sections for $e\,p \to e^\prime \,h\,X$ and  $p\, p\to \ell\, \ell^\prime\,X$, we are able to assess the role of the shape functions in the TMD formalism for quarkonium production. This will have important implications for a recent suggestion to extract poorly known CO LDMEs from a comparison between quarkonium production and open heavy quark pair production in SIDIS at the EIC \cite{Bacchetta:2018ivt}.

The paper is organized as follows. In Section~\ref{sec:h-l} we define the variables  that are adopted in our calculation. Parametrizing the cross section in terms of different structure functions, we compute the cross section in the collinear framework after which we take the small-$q_\sT$ limit. Section~\ref{sec:l-h} is devoted to the computation of the cross section in the TMD regime, under the approximation that the $J/\psi$ meson is collinear with the outgoing heavy-quark pair. The large-$q_\sT$ limit of the result is then taken and compared with the small-$q_\sT$ limit of the collinear calculation. In Section~\ref{sec:shape}, both results are shown to match after including the smearing of the transverse momentum of the quarkonium in its hadronization, which is encoded in the appropriate shape functions. Conclusions are given in Section~\ref{sec:conc}. Finally, details on the reference frames and the expansion of the momentum conserving delta function in the small-$q_\sT$ limit can be found in the appendices.

\section{From high to intermediate transverse momentum}
\label{sec:h-l}
In this section collinear factorization and NRQCD are adopted for the description of the process 
\begin{equation}
e(\ell) + p(P) \to e(\ell^{\prime}) + J/\psi(P_\psi) + X\,,
\end{equation}
where all the particles are unpolarized and their four-momenta are given within brackets. This reaction is described by the conventional variables
\begin{equation}
\xB = \frac{Q^2}{2P\cdot q}\,, \qquad y = \frac{P\cdot q}{P\cdot \ell}\, , \qquad z = \frac{P\cdot P_\psi}{P\cdot q}\,,
\end{equation}
with $q\equiv \ell-\ell^\prime$ and $q^2 = -Q^2$.  We denote by $M_p$ and $M_\psi$ the masses of the proton and the $J/\psi$ meson, respectively. 
In the one-photon exchange approximation, at leading order (LO) in perturbative QCD, i.e.\ $\alpha_s^2$, the transverse momentum of the $J/\psi$ is due to parton emission in the hard scattering process
\begin{align}
\gamma^* (q) + a(p_a) \to c \overline c[n](P_\psi) + a(p_a^\prime)\,, 
\end{align}
where parton $a$  can be either a gluon, a quark or an antiquark, and the charm-anticharm quark pair is produced in a specific Fock state $n = \, ^{2S+1\!}L_J^{[c]}$. 
We also introduce the scaling variables
\begin{equation}
\hat{x} = \frac{Q^2}{2p_a\cdot q}\,,\qquad \hat z = \frac{p_a\cdot P_\psi}{p_a\cdot q}\,. 
\end{equation}
If we neglect the proton mass and any smearing effects both in the initial and in the final state, we can take
\begin{align}
p_a = \xi P \,,
\end{align}
and therefore
\begin{align}
\hat x = \frac{\xB}{\xi}\,,\qquad   \hat z = z\,,
\end{align}
which implies
\begin{align}
\hat x p_a = \xB P\, \qquad {\rm and}\qquad \hat x \ge \xB\,.
\end{align}

\begin{figure}[t]
\begin{centering}
\subfloat[]{\includegraphics[trim={8cm 22.8cm 8cm 3cm},clip,scale=1]{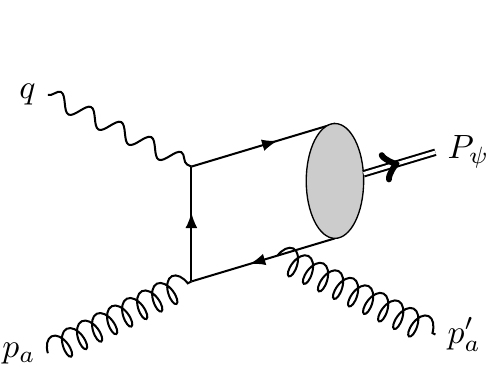} } \hspace*{2cm}
\subfloat[]{\includegraphics[trim={5.5cm 22.8cm 6.5cm 3cm},clip,scale=1]{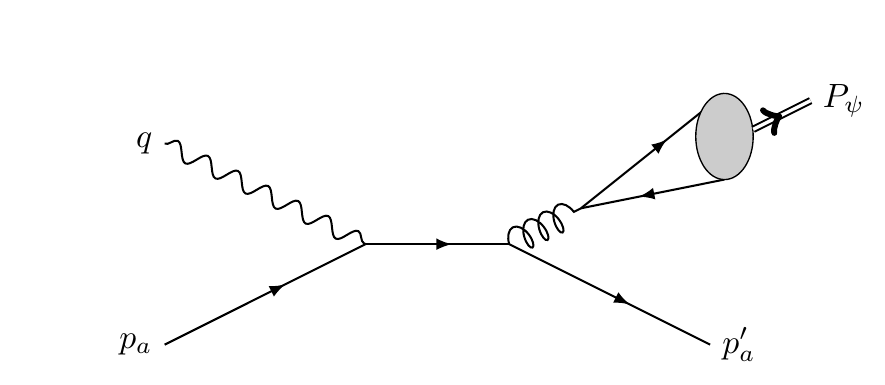}}

\hspace*{-2.cm}\subfloat[]{\includegraphics[trim={7.cm 21.8cm 8cm 3cm},clip,scale=1]{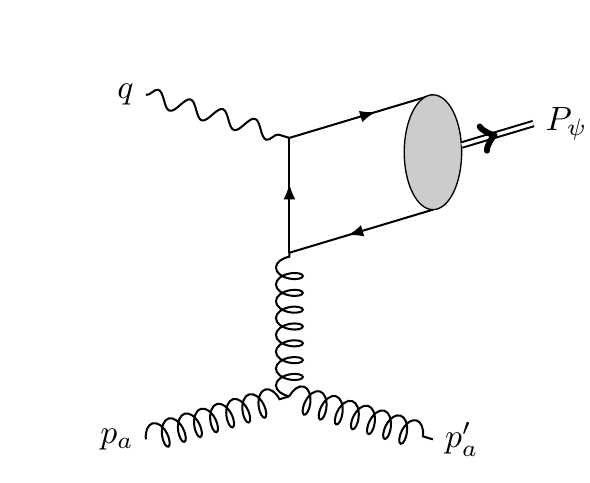}} \hspace*{3.cm}
\subfloat[]{\includegraphics[trim={7.cm 21.8cm 8cm 3cm},clip,scale=1]{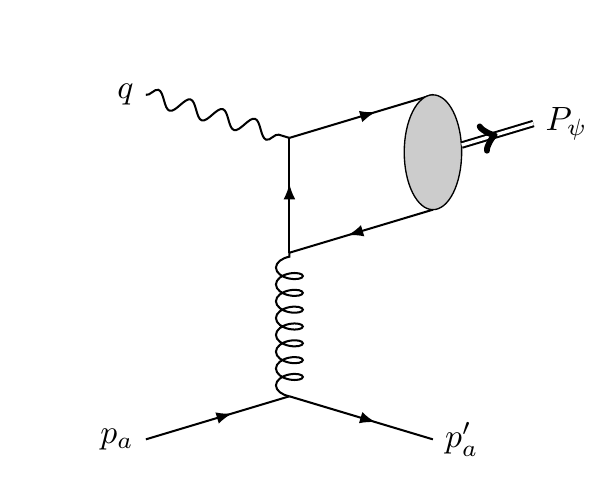}}
\par\end{centering}
\caption{Representative diagrams for the partonic process $\gamma^* (q) \,+ \, a(p_a)\to {J/\psi} (P_\psi) \, + \, a(p_a^\prime)$, with $a = g, q, \bar q$ at the order $\alpha_s^2$. The only diagrams contributing to the CS production mechanism are of the type (a), and there are six of them. There are two diagrams for each type (b), (c), (d). The dominant diagrams in the small-$q_\sT$ limit are those of type (c) and (d).}
\label{fig:fd}
\end{figure}

In a frame where the longitudinal directions are fixed by the proton and the photon,  the cross section for the process under study can be written as follows, 
\begin{align}
\frac{\d\sigma}{\d y\, \d \xB \,\d z\, \d \bm P^2_{\psi\perp}\d \phi_\psi } & = \frac{1}{64}\,\frac{1}{(2\pi)^4} \, y \sum_n\int_{\xB}^{\hat x_{\rm max}}
 \frac{\d \hat x}{\hat x}\int_z^1  \frac{\d \hat z}{\hat z}\, \delta \left ( \frac{\bm q_\sT^2}{Q^2}  + \frac{1-\hat z}{\hat z^2 }\, \frac{M_\psi^2}{Q^2} -\frac{(1-\hat x)(1- \hat z) }{\hat x \hat z}  \right ) \nonumber \\
& \qquad \times \sum_a\left [ \frac{1}{Q^6} \,  f_1^a \left (\frac{\xB}{ \hat x} \, ,\mu^2\right )\, L^{\mu\nu}\,H^{a [n]}_{\mu}\,H^{a [n]*}_{\nu}  \, \langle 0 \vert {\cal O} (n)\vert 0 \rangle \right ] \frac{1}{z} \, \delta (\hat z-z) \,,
\label{eq:cs-start}
\end{align}
where  $\bm P_{\psi\perp}$ and $\phi_\psi$ are the transverse momentum and the azimuthal angle of the final $J/\psi$ meson, defined with respect to the lepton plane according to the conventions of Ref.~\cite{Bacchetta:2004jz}, and $\bm q_\sT^2 = \bm P_{\psi \perp}^2/\hat z^2$ (see Appendix~\ref{sec:frames} for details). Moreover,  $H^{a [n]}_{\mu}$ is the amplitude for the hard scattering subprocess $\gamma^* a\to c \bar{c}[n]\,a$, with $a=g,q,\bar{q}$. The corresponding Feynman diagrams, at the perturbative order $\alpha_s^2$, are depicted in Fig.~\ref{fig:fd}. The Fock states included in the calculation are $n=\,^3\!S_1^{[1]}$, $^1\!S_0^{[8]}$,  $^3\!S_1^{[8]}$, $^3\!P_J^{[8]}$, with $J = 0,1,2$~\cite{Kniehl:2001tk,Sun:2017nly,Sun:2017wxk,Zhang:2019ecf}. Furthermore, in Eq.~\eqref{eq:cs-start}, $f_1^a$ is the unpolarized PDF, which depends on the light-cone momentum fraction $\xi= \xB/\hat x$ of parton $a$ and on the a priori arbitrary hard factorization scale $\mu$. Even if not explicitly indicated, the hard scattering amplitudes  $H^{a [n]}_{\nu}$ depend on the scale $\mu$ as well.

The leptonic tensor $L^{\mu\nu}$ can be written as \cite{Bacchetta:2006tn}
\begin{align}
L^{\mu\nu} & =  e^2  \big [ - g^{\mu\nu} Q^2 + 2 (\ell^\mu \ell^{\prime \nu} + \ell^\nu \ell^{\prime \mu}) \big ]\nonumber \\
& = e^2\frac{Q^2}{y^2}\, \bigg  \{ -[ 1+(1-y)^2 ]\, g_\perp^{\mu\nu} \,+ \, 4 (1-y)\,  \epsilon^\mu_L \epsilon^\nu_L  \,+\, 4 (1-y) 
\left ( \hat \ell_\perp^\mu  \hat \ell_\perp ^\nu   + \frac{1}{2}\, g_\perp^{\mu\nu}\right )  \nonumber \\
& \qquad \qquad\qquad \qquad  + 2 (2-y)\, \sqrt{1-y} \, ( \epsilon_L ^{\mu}\,  \hat \ell_\perp^{\nu}  +  \epsilon_L ^{\nu}\,  \hat  \ell_\perp^{\mu} ) \bigg \} \,,
\end{align}
where the second equality can be obtained from the first one by replacing the expression for $\ell^\mu$ in Eq.~\eqref{eq:cs-pT}, and where 
the transverse projector $g_\perp^{\mu\nu}$ is given by 
\begin{align}
g_\perp^{\mu\nu} & \equiv g^{\mu\nu} - \kappa_+^\mu \kappa_-^\nu - \kappa_-^\mu \kappa_+^\nu = g^{\mu\nu} - \frac{1}{P\cdot q} \, (P^\mu q^\nu + P^\nu q^\mu) - \frac{Q^2}{(P\cdot q)^2}\,P^\mu P^\nu\,.
\end{align}
Furthermore, we have introduced the longitudinal polarization vector of the exchanged virtual photon, 
\begin{align}
\epsilon_L^\mu (q)= \frac{1}{Q}\left (  q^\mu + \frac{Q^2}{P\cdot q} \, P^\mu \right ) \,,
\end{align}
which fulfills the relations $\epsilon^2_L (q)= 1$ and $\epsilon_L^\mu(q)\, q_\mu = 0$. 
From Eq.~\eqref{eq:qT-PhT} it follows that the cross section differential in $\bm q_\sT^2$ can be obtained by simply multiplying Eq.~\eqref{eq:cs-start} by a factor $z^2$\,,
\begin{align}
\frac{\d\sigma}{\d y\, \d \xB \,\d z\, \d \bm q_\sT^2 \,\d \phi_\psi } & = \frac{1}{64}\,\frac{1}{(2\pi)^4} \,{y z} \sum_n\int_{\xB}^{\hat x_{\rm max}}
 \frac{\d \hat x}{\hat x}\int_z^1  \frac{\d \hat z}{\hat z}\, \delta \left ( \frac{\bm q_\sT^2}{Q^2}  + \frac{1-\hat z}{\hat z^2 }\, \frac{M_\psi^2}{Q^2} -\frac{(1-\hat x)(1- \hat z) }{\hat x \hat z}  \right )  \nonumber \\
& \qquad \times \sum_a\left [ \frac{1}{Q^6} \,  f_1^a \left (\frac{\xB}{ \hat x} \,,\mu^2\right ) \,L^{\mu\nu}  H^{a[n]}_{\mu}\,H^{a[n]*}_{\nu}   \, \langle 0 \vert {\cal O} (n)\vert 0 \rangle \right ]  \delta (\hat z -z) \,.
\label{eq:cs}
\end{align}
Along the same lines of Ref.~\cite{Bacchetta:2006tn}, the final result can be expressed in terms of four independent structure functions:
\begin{align}
\frac{\d\sigma}{\d y\, \d \xB \,\d z\, \d \bm q_\sT^2 \,\d \phi_\psi } & =\frac{\alpha^2}{y Q^2} \bigg\{ [1+(1-y)^2]\, F_{UU,T} + 4 {(1-y)}\, F_{UU,L}  \,\nonumber \\
& \qquad \qquad  + (2-y) \sqrt{1-y} \, \cos\phi_\psi\, F_{UU}^{\cos\phi_\psi}  \,+ (1-y) \cos2\phi_\psi\, F_{UU}^{\cos2\phi_\psi} \bigg\} \,,
\end{align}
where the first and second subscripts of the structure functions $F$ denote the polarization of the initial electron and proton, respectively, while the third one, when present, specifies the polarization of the exchanged virtual photon.

The full expressions of the structure functions can be found, for example, in Ref.~\cite{Sun:2017nly}. Here, we are mainly interested in the behaviour of the cross section in the  $\bm q_\sT^2\ll Q^2$ region. This can be obtained from Eq.~\eqref{eq:cs}, replacing the Dirac delta with its expansion in the small-$q_\sT$ limit derived in Appendix~\ref{sec:delta-exp}, namely
\begin{align}
\delta \left (\frac{\bm q_\sT^2}{Q^2} + \frac{1-\hat z}{\hat z^2} \, \frac{M_\psi^2}{Q^2}- \frac{(1-\hat x)(1- \hat z)}{\hat x \hat z} \right ) &  = \hat x_{\rm max} \left \{\frac{\hat x^\prime }{(1- \hat  x^\prime)_+} \, \delta (1-\hat z) \, +
\,   \frac{Q^2 + M^2_\psi}{Q^2 + M_\psi^2/\hat z}   \,  \frac{\hat z}{(1-\hat z)_+} \ \delta \left (1 -\hat x^\prime \right )  \right .\nonumber \\
&\qquad  \left .+ \,  \delta (1- \hat x^\prime) \delta (1-\hat z)  \ln \bigg(\frac{Q^2+M_\psi^2}{\bm q_\sT^2}   \bigg) \right \}\,,
\end{align}
where 
\begin{align}
\hat x_{\rm max} = \frac{Q^2}{Q^2 + M_\psi^2}\,,  \qquad  \hat x^\prime = \frac{\hat x}{\hat x_{\rm max}}\,.
\end{align}
By using the relations
\begin{align}
\frac{1}{1-\hat z} = \frac{1}{\bm q_\sT^2}\, \frac{Q^2(1-\hat x^\prime) \hat z + M_\psi^2(\hat z - \hat x^\prime) }{\hat x^\prime \hat z^2 }\,,
\end{align}
and
\begin{align}
\frac{1}{1-\hat x^\prime} = \frac{\hat z (1-\hat z)}{\hat x^\prime}\, \frac{Q^2+M^2_\psi }{\hat z^2 \bm q_\sT^2 + (1-\hat z)^2M^2_\psi}\,,
\end{align}
we obtain the leading power behavior of the structure functions
\begin{align}
F_{UU,T}  & = \sigma_{UU,T} \left [  L\left (  \frac{Q^2+M_\psi^2}{\bm q_\sT^2} \right ) f_1^g(x,\mu^2) + \left (P_{gg} \otimes  f_1^g\,+  P_{gi} \otimes f_1^i \right ) (x,\mu^2) \right ]\,,\nonumber \\
F_{UU,L}  & = \sigma_{UU,L} \,\left  [  L\left (  \frac{Q^2+M_\psi^2}{\bm q_\sT^2} \right ) f_1^g(x,\mu^2) + (P_{gg} \otimes  f_1^g\,+\,  P_{gi} \otimes f_1^i)  (x,\mu^2)  \right ]\,,\nonumber \\
F_{UU}^{\cos2\phi_{\psi}}   & =  \sigma_{UU}^{\cos 2\phi_\psi}\, \left [ (\delta P_{gg}\otimes f_1^g )(x,\mu^2) +  (\delta P_{gi}\otimes f_1^i )(x,\mu^2) \right ]\,,
\label{eq:str-funct-coll}
\end{align}
where a sum over $i = q, \bar{q}$ is understood. These  results are valid up to corrections of the order of  ${\cal O} ( \Lambda_{\rm QCD} / \vert \bm q_\sT\vert)$ and ${\cal O} (\vert \bm q_\sT\vert/Q)$.  The structure function  $F_{UU}^{\cos\phi_{\psi}}$ is suppressed by a factor $\vert \bm q_\sT\vert/Q$ with respect to the other ones and will not be considered in the following.  In Eq.~\eqref{eq:str-funct-coll}, we have defined
\begin{align}
x \equiv \frac{\xB}{\hat x_{\rm max}} = \xB\, \left (  1+ \frac{M^2_\psi}{Q^2} \right )  \,,
\end{align}
and
\begin{align}
L \left ( \frac{Q^2+ M^2_\psi}{\bm q_\sT^2} \right ) \equiv 2 C_A\, \ln\left ( \frac{Q^2+ M^2_\psi}{\bm q_\sT^2} \right )- \frac{11C_A-4n_fT_R}{6}\, ,
\label{eq:L-glu}
\end{align}
where $n_f$ refers to the number of active flavors,  $T_R=1/2$, $C_A= N_c$, with $N_c$ being the number of colors. The symbol $\otimes$ denotes a convolution in the longitudinal momentum fractions:
\begin{align}
(P\otimes f)(x,\mu^2) = \int_x^1\frac{\d\hat x}{\hat x}\, P \left ( \hat x, \mu^2 \right )  f \left ( \frac{x}{\hat x} , \mu^2\right ) \,.
\end{align}
The well-known LO unpolarized splitting functions read
\begin{align}
P_{gg} (\hat x) & = 2 C_A \left [ \frac{\hat x}{(1-\hat x)_+} + \frac{1-\hat x}{\hat x} + \hat x (1-\hat x) \right ] + \delta (1-\hat x) \, \frac{11C_A-4n_f T_R}{6}\,,\nonumber \\
P_{gq} (\hat x)& = P_{g\bar q}  (\hat x) =C_F \, \frac{1+ (1-\hat x)^2}{\hat x}\,,
\end{align}
while the splitting functions of an unpolarized parton into a linearly polarized gluon are~\cite{Sun:2011iw,Catani:2010pd} 
\begin{align}
\delta P_{gg}(\hat x) & = C_A\, \frac{1-\hat x}{\hat x}\,, \nonumber \\
\delta P_{gq}(\hat x) & = \delta P_{g\bar q}(\hat x)= C_F\, \frac{1-\hat x}{\hat x}\,, 
\end{align}
with $C_F = (N_c^2-1)/2 N_c$. The plus-prescription on the singular parts of the splitting functions is defined, as usual,  such that the integral of a sufficiently smooth distribution $G$ is given by 
\begin{align}
\int_z^1\d y \,  \frac{G(y)}{(1-y)_+} = \int_z^1\d y\,  \frac{G(y)-G(1)}{1-y} - G(1) \ln \left ( \frac{1}{1-z} \right )
\label{eq:plus}
\end{align}
and 
\begin{align}
\frac{1}{(1-y)_+} = \frac{1}{1-y}\qquad {\rm for}~~ 0 \le y < 1\,. 
\end{align}
Assuming the validity of the common heavy-quark spin symmetry relations~\cite{Bodwin:1994jh}
\begin{align}
\langle 0 \vert{\cal  O} (^3P_J^{[8]})\vert  0 \rangle = (2J+1) \,\langle 0 \vert {\cal O} (^3P_0^{[8]})\vert 0 \rangle\,,
\end{align}
the cross sections for the partonic processes $\gamma^*g\to c \overline c [n]$  in Eq.~\eqref{eq:str-funct-coll} read
\begin{align}
\sigma_{UU,T} & = \frac{ \alpha_s^2e_c^2}{M_\psi (M_\psi^2 + Q^2) \bm q_\sT^2}\,
\left [ \langle 0 \vert {\cal O}(^1S_0^{[8]})\vert  0\rangle  \, +\,4\,\frac{(7 M_\psi^4 + 2 M_\psi^2 Q^2 + 3Q^4)}{M_\psi^2(M_\psi^2 + Q^2)^2} \langle 0 \vert {\cal O}(^3P_0^{[8]})\vert  0\rangle  \right ] \,,\nonumber \\
\sigma_{UU,L} & =  \frac{ \alpha_s^2e_c^2}{M_\psi (M_\psi^2 + Q^2) \bm q_\sT^2}\, \left [16\, \frac{ Q^2}{(M_\psi^2 + Q^2)^2}\, \langle 0 \vert {\cal O}(^3P_0^{[8]})\vert  0\rangle  \right ] \,,\nonumber \\
\sigma_{UU}^{\cos 2\phi_\psi} & = \frac{4 \alpha_s^2 e^2_c}{M_\psi (M^2_\psi + Q^2)\bm q_\sT^2}\left [ - \langle 0 \vert {\cal O} (^1S_0^{[8]})\vert  0\rangle \, + \, 4\, \frac{3M^2_\psi -Q^2}{M^2_\psi (M^2_\psi + Q^2)} \,\langle 0 \vert {\cal O}(^3P_0^{[8]})\vert  0\rangle \right ] \, , 
\label{eq:cs-small-qT}
\end{align}
where $e_c$ is the electric charge of the charm quark in units of the proton charge.  

We point out that  in the present study at the order $\alpha_s^2$ we include the CS and those subleading CO contributions that are relatively suppressed by a factor of $v^4$. Nevertheless, we cannot claim that our result will describe the bulk of the SIDIS cross section. Indeed, higher order corrections in $\alpha_s$ but with an enhanced $v$ scaling, possibily with a different $\bm q_\sT^2$-behavior, can be relevant and need to be investigated in future studies. Moreover, we note that the partonic subprocesses contributing to the cross sections in the small-$q_\sT$ limit given in Eq.~\eqref{eq:cs-small-qT} are only the  $n =~ ^1\!S_0^{[8]}$, $^3\!P_J^{[8]}$ ones, which correspond to $\hat t$-channel Feynman diagrams of the types (c) and (d) in Fig.~\ref{fig:fd}.  The other partonic subprocesses, namely the gluon induced $^3\!S_1^{[1,8]}$ channels and the quark-induced $^1\!S_0^{[8]}$ channel depicted in  Figs.~\ref{fig:fd}~(a)-(b), are suppressed and vanish as $\bm q_\sT^2 \to 0$. Hence they are not relevant for our study  of the matching of the collinear and TMD results and will not be considered in the following.

Coming back to the structure functions $F_{UU,T}$ and $F_{UU,L}$ in Eq.~\eqref{eq:str-funct-coll}, we note that  they exhibit  logarithmic (collinear) singularities as $q_\sT\to 0$. Their behavior is similar to the analogous structure functions for light hadron production in SIDIS discussed in Ref.~\cite{Bacchetta:2008xw}, where the dominant underlying partonic process is $\gamma^* q\to q$. There are some differences though. In the latter case, the logarithmic term $L$ is given by $C_F (2 \ln Q^2/\bm q_\sT^2 - 3)$ instead of Eq.~\eqref{eq:L-glu}. The color factor $C_F$  clearly corresponds to a quark initiated process, while $C_A$ corresponds to a gluon initiated one. The two different finite terms originate from the virtual corrections to the splitting functions $P_{qq}$ and $P_{gg}$, respectively. Moreover, in light hadron production extra terms  appear, containing convolutions of FFs with the $P_{qq}$ and $P_{gq}$ splitting functions, which cannot be present in our calculation for quarkonium production within the NRQCD framework. We also point out that the structure function $F_{UU}^{\cos2\phi_\psi}$ does not contain any large logarithm in the region $\bm q_\sT^2 \ll Q^2$, whereas the corresponding observable for light hadron production diverges logarithmically and is suppressed by an overall factor $\bm q_\sT^2/Q^2$. Finally, the appearance of a logarithm $\ln (Q^2+M_\psi^2)/\bm q_\sT^2$, instead of $\ln Q^2/\bm q_\sT^2$, suggests $Q^2+M_\psi^2$ as the natural choice for the hard scale in the process under study. 

\section{From small to intermediate transverse momentum}
\label{sec:l-h}

The process  $\gamma^*g\to c \overline c[n]$ has been calculated in Ref.~\cite{Bacchetta:2018ivt} within the TMD framework, taking into account the intrinsic transverse momentum effects of the gluons inside the proton. If we neglect smearing effects in the final state, {\it i.e.}\ if we assume that the final $J/\psi$ meson is  collinear to the $c\bar c$ pair originally produced in the hard scattering process, the cross section can be cast in the following form
\begin{align}
\frac{\d\sigma}{\d y\, \d \xB \,\d z\, \d  \bm q_\sT^2 \,\d \phi_\psi } & = \, \frac{\alpha^2}{y Q^2} \,  \left \{ [ {1+(1-y)^2}]\, 
{\cal F}_{UU,T}   \,+\, 4(1-y)  \, {\cal F}_{UU,L}   \,+\, (1-y)\, \cos2\phi_\psi\,{\cal F}_{UU}^{\cos2\phi_\psi}  \right \} \delta(1-z)
\end{align}
with 
\begin{align}
{\cal F}_{UU,T}  & = \frac{2\pi^2\alpha_s e^2_c}{M_\psi (M^2_\psi+ Q^2)}\, \left [ \langle 0 \vert {\cal O} (^1S_0^{[8]})\vert  0\rangle  \,  +\,4\,\frac{(7 M_\psi^4 + 2 M_\psi^2 Q^2 + 3Q^4)}{M_\psi^2(M_\psi^2 + Q^2)^2} \langle 0 \vert {\cal O} (^3P_0^{[8]})\vert  0\rangle  \right ] f_1^g(x, p_\sT^2) \bigg \vert_{\bm p_\sT = \bm q_\sT}\,, \nonumber \\
{\cal F}_{UU,L}  & =   \frac{2\pi^2\alpha_s e^2_c}{M_\psi (M^2_\psi + Q^2)} \,\frac{16\, Q^2}{ (M^2_\psi + Q^2)^2}\, \langle 0 \vert {\cal O} (^3P_0^{[8]}) \,\vert  0\rangle \,f_1^g(x,  p_\sT^2) \bigg \vert_{\bm p_\sT = \bm q_\sT}\, , \nonumber \\
{\cal F}_{UU}^{\cos2\phi_\psi}  & =  \frac{2\pi^2\alpha_s e^2_c}{M_\psi (M^2_\psi + Q^2)}\left [ - \langle 0 \vert {\cal O} (^1S_0^{[8]})\vert  0\rangle \, + \, 4\, \frac{3M^2_\psi -Q^2}{M^2_\psi (M^2_\psi + Q^2)} \,\langle 0 \vert {\cal O}(^3P_0^{[8]})\vert  0\rangle \right ]\frac{\bm p_\sT^2}{M_p^2}\,   h_1^{\perp\,g}(x, p_\sT^2) \bigg \vert_{\bm p_\sT = \bm q_\sT} \,,
\label{eq:SF-TMD}
\end{align}
where $f_1^g$ and $h_1^{\perp\, g}$ are, respectively,  the unpolarized and linearly polarized gluon TMDs inside an unpolarized proton~\cite{Mulders:2000sh,Meissner:2007rx,Boer:2016xqr,Echevarria:2015uaa,Gutierrez-Reyes:2019rug,Luo:2019bmw}. 

We note that, beyond the parton model approximation, for those processes where TMD factorization is valid, soft gluon radiation to all orders is included into an exponential Sudakov factor, which can be split and its parts absorbed into the TMD PDFs and FFs involved in the reaction, whereas the remaining perturbative corrections are collected into a hard factor ${\cal H}$. As a consequence of the regularization of their ultraviolet and rapidity divergences, TMDs depend on two different scales, not explicitly shown in the above equations.   In the following we will take these two scales to be equal to each other and denote them by $\mu$, to be identified with a typical hard scale of the process. 
 
TMDs can be calculated perturbatively in the limit $ \vert \bm p_\sT \vert \gg\Lambda_{\rm QCD}$.  This can be better achieved in the impact parameter space. To this aim, we focus first on $f_1^g(x, p_\sT^2 )$ and its Fourier transform, which  is defined as
\begin{align}
\widehat f_1^g (x,\bm b_\sT^2; \mu^2 ) \equiv \frac{1}{2\pi}\int \d^2 \bm p_\sT \,e^{i \bm b_\sT\cdot \bm p_\sT}f_1^g(x,  \bm p_\sT^2;\mu^2) = \int_0^\infty \d \vert \bm p_\sT \vert \,\vert \bm p_\sT\vert \, J_0(\vert \bm b_\sT \vert \vert \bm p_\sT\vert ) f_1^g(x, \bm p_\sT^2; \mu^2)\,, 
\end{align}
with $J_n$ being the Bessel function of the first kind of order $n$. The perturbative part of the gluon TMD, valid in the limit $\vert \bm b_\sT\vert \ll 1/\Lambda_{\rm QCD}$, reads~\cite{Collins:2011zzd}
\begin{align}
\widehat f_1^g(x,\bm b_\sT^2; \mu^2)=\frac{1}{2\pi} \sum_{a = q, \bar q, g} (C_{g/a} \otimes f_1^a) (x, \mu_b^2) \, e^{-\frac{1}{2}S_A(\bm b_\sT^2, \mu^2)}\, , 
\label{eq:f1g-FT}
\end{align}
where $f_1^a(x,\mu^2)$  are the collinear parton distribution functions for a specific (anti)quark flavor or a gluon $a$, and $\mu_b=b_0/\vert \bm b_\sT\vert$ with $b_0 = 2e^{-\gamma_E} \approx 1.123$. 
The coefficient functions $C_{g/a}$ and the perturbative Sudakov exponent $S_A$, which resums large logarithms of the type $\ln (b_\sT \mu)$, are calculable in perturbative QCD. 
The coefficient functions can be expanded in powers of $\alpha_s$ as follows
\begin{align}
C_{g/a}(x, \mu_b) = \delta_{ga}\, \delta(1-x) + \sum_{k=1}^\infty C^{(k)}_{g/a}(x) \left (\frac{\alpha_s(\mu_b)}{\pi} \right )^k\,,
\end{align}
while the perturbative Sudakov factor at LO reads 
\begin{align}
S_A(\bm b_\sT^2, \mu^2) & = \frac{C_A}{\pi}\int_{\mu_b^2}^{\mu^2}\frac{\d\zeta^2}{\zeta^2}\, \alpha_s(\zeta)\left ( \ln \frac{\mu^2}{\zeta^2} - \frac{11- 2 n_f/C_A}{6} \right ) = \frac{C_A}{\pi}\, \alpha_s \left (\frac{1}{2}  \ln^2 \frac{\mu^2}{\mu_b^2}  -  \frac{11- 2 n_f/C_A}{6}\, \ln \frac{\mu^2}{\mu_b^2}  \right ) \,,
\label{eq:Sud}
\end{align}
where in the last equation, valid to the order $\alpha_s$, the running of the QCD coupling constant is neglected since it enters at the order $\alpha_s^2$~\cite{Bacchetta:2008xw,Bacchetta:2019qkv}. For this reason, from now on the argument of the coupling constant is not shown.

By substituting Eq.~\eqref{eq:Sud} into Eq.~\eqref{eq:f1g-FT},  we find, in the small-$b_\sT$ limit, namely $b_\sT \ll 1/\Lambda_{\rm QCD}$, and at LO in $\alpha_s$,
\begin{align} 
\widehat f_1^g(x,\bm b_\sT^2; \mu^2)=\frac{1}{2\pi} \left \{ f_1^g(x, \mu_b^2) - \frac{\alpha_s}{2\pi} \,\left [  \left (\frac{C_A}{2}\, \ln^2 \frac{\mu^2}{\mu_b^2}  -  \frac{11C_A- 2 n_f}{6}\, \ln \frac{\mu^2}{\mu_b^2}  \right )    f_1^g(x, \mu_b^2) \,-\,  2 \sum_a \left (  C_{g/a}^{(1)} \otimes f_1^a\right )(x, \mu_b^2)  \right ]  \right \}\,.
\label{eq:f1-FT}
\end{align}
The explicit expressions of the hard coefficients $C_{g/a}^{(1)}$, not relevant for our analysis, can be found in Eqs.~(A33) and (A35) of Ref.~\cite{Echevarria:2015uaa}.
Using the DGLAP equations we can evolve $f_1^g$ from a scale $\mu$ down to another scale $\mu_b < \mu$  and obtain
\begin{align}
f_1^g(x,\mu_b^2) = f_1^g(x, \mu^2) - \frac{\alpha_s}{2\pi} \, (P_{gg} \otimes f_1^g  \,+\,  P_{gi}\otimes f_1^i) (x, \mu^2) \ln \frac{\mu^2}{\mu_b^2} + {\cal O}(\alpha_s^2)\,,
\end{align}
where a sum over $i = q, \bar q$ is understood. By substituting the above expression into Eq.~\eqref{eq:f1-FT} we get
\begin{align}
\widehat f_1^g(x,\bm b_\sT^2; \mu^2)&= \frac{1}{2 \pi}  \bigg \{ f_1^g(x, \mu^2) - \frac{\alpha_s}{2\pi} \,\bigg [ \bigg ( \frac{ C_A}{2}\,\ln^2 \frac{\mu^2}{\mu_b^2}  -  \frac{11C_A- 2 n_f }{6}\, \ln \frac{\mu^2}{\mu_b^2}  \bigg )    f_1^g(x, \mu^2)  
 \nonumber \\
& \qquad  +  (P_{gg} \otimes f_1^g  \,+\,  P_{gi}\otimes f_1^i) (x,\mu^2) \ln \frac{\mu^2}{\mu_b^2}\,-\,  2 \sum_a \left (  C_{g/a}^{(1)} \otimes f_1^a\right )(x, \mu^2) \bigg ]  \bigg \} \,.
\label{eq:f1-FT-final}
\end{align}
Transforming back to momentum space, we find the transverse momentum distribution in the region $\vert \bm p_\sT \vert \gg \Lambda_{\rm QCD}$, 
\begin{align}
f_1^g(x, \bm p_\sT^2; \mu^2) & = \frac{1}{2\pi}\int \d^2 \bm b_\sT\, e^{-i \bm b_\sT \cdot \bm p_\sT}\,\widehat  f_1^g(x, \bm b_\sT^2;\mu^2) \, \nonumber \\
& = \frac{\alpha_s}{2\pi^2 \bm p_\sT^2}\, \,\left [ \left (C_A  \ln \frac{\mu^2}{ p_\sT^2}  -  \frac{11 C_A- 2 n_f }{6}\,    \right )    f_1^g(x, \mu^2) \,+\,(P_{gg} \otimes f_1^g  \,+\,  P_{gi}\otimes f_1^i) (x, \mu^2) \right ]  \,,
\label{eq:TMD-f1}
\end{align}
where we have used the following integrals
\begin{align}
\int \d^2 \bm b_\sT \,e^{-i \bm b_\sT \cdot \bm q_\sT}\ln^2 \frac{\mu^2}{\mu_b^2} = - \frac{8 \pi}{\bm q_\sT^2}\,\ln \frac{\mu^2}{ \bm q_\sT^2}\,, \qquad \int \d^2 \bm b_\sT \,e^{-i \bm b_\sT \cdot \bm q_\sT}\ln \frac{\mu^2}{\mu_b^2} = - \frac{4 \pi}{ \bm q_\sT^2}\,.
\end{align}
Moreover, we note that the term at the order $\alpha_s^0$  and the coefficients $C^{(1)}_{g/a}$ do not appear in the final result because  they are independent of $b_\sT$ and hence give a contribution to $f_1^g$ proportional to $\delta^2(\bm p_\sT)$.  Since we require $\vert \bm p_\sT \vert \gg \Lambda_{\rm QCD}$, such terms will be discarded. For the same reason, the unpolarized gluon distribution in the  high-$p_\sT$ region does not depend on any nonperturbative model~\cite{Bacchetta:2008xw,Bacchetta:2019qkv}.

By substituting the expression for $f_1^g(x, p_\sT^2)$ given in Eq.~\eqref{eq:TMD-f1}, evolved to the scale $\mu^2 = Q^2+M_\psi^2$, into Eq.~\eqref{eq:SF-TMD}, we find that the  TMD structure functions ${\cal F}_{UU,T}$ and ${\cal F}_{UU,L}$  do not  exactly match the corresponding collinear ones in the small-$q_\sT$ limit given in Eq.~\eqref{eq:str-funct-coll}. In the intermediate region $\Lambda_{\rm QCD} \ll \vert \bm q_\sT \vert  \ll Q$, we get
\begin{align}
{\cal F}_{UU,T}  & =F_{UU,T} - \sigma_{UU,T}\,C_A \ln \left ( \frac{Q^2+M^2_\psi}{\bm q_\sT^2} \right )   \,,\nonumber \\
{\cal F}_{UU,L} & = F_{UU,L} - \sigma_{UU,L}\,C_A \ln \left ( \frac{Q^2+M^2_\psi}{\bm q_\sT^2}  \right )  \,.
\end{align}
This suggests that one needs to include smearing effects in the final state as well, through the inclusion of a suitable shape function~\cite{Echevarria:2019ynx,Fleming:2019pzj}, to be convoluted with the TMD in momentum space. Imposing the validity of the matching will give us the LO expression of the shape function, as we shall see in the next section. 

We now turn to the polarized gluon distribution $h_1^{\perp\,g}(x, \bm  p_\sT^2)$  and the structure function ${\cal F}_{UU}^{\cos2\phi_\psi}$.  The perturbative tail  of $h_1^{\perp\,g}$ can be calculated along the same lines of $f_1^g$, with the important difference that its expansion in powers of the QCD coupling constant starts at ${\cal O}(\alpha_s)$. As a consequence, at leading order $h_1^{\perp\,g}$ does not depend on the soft factor and its expression can  be taken directly from Ref.~\cite{Sun:2011iw}, where it has been obtained within the traditional Collins-Soper-Sterman resummation framework~\cite{Collins:1984kg}, in which the soft factor is not part of the definition of the TMD. The result  at LO in terms of the unpolarized collinear PDFs $f_1^{a}(x,\mu^2)$ reads
\begin{align}
 \frac{ \bm p_\sT^2}{2 M_p^2}\,h_1^{\perp g } (x,  \bm p_\sT^2; \mu^2) 
 & =  \frac{\alpha_s}{\pi^2} \,  \frac{1}{\bm p_\sT^2}\, \left [ C_A \int_x^1 \frac{\d  \hat x}{\hat x} \, \left  ( \frac{\hat x}{x} -1  \right )\, f_1^{g }(\hat x, \mu^2)\, + \, C_F \sum_{i = q, \bar q} \, \int_x^1 \frac{\d  \hat x}{\hat x} \, \left  ( \frac{\hat x}{x} -1  \right )\, f_1^{i }(\hat x, \mu^2 )
 \right ]\,.
 \label{eq:h1p}
\end{align}
The above expression, together with Eq.~\eqref{eq:SF-TMD}, leads to 
\begin{align}
{\cal F}_{UU}^{\cos2\phi_\psi}  & =  \frac{4 \alpha_s^2 e^2_c}{M_\psi (M^2_\psi + Q^2)\bm q_\sT^2}\left [ - \langle 0 \vert {\cal O} (^1S_0^{[8]})\vert  0\rangle \, + \, 4\, \frac{3M^2_\psi -Q^2}{M^2_\psi (M^2_\psi + Q^2)} \,\langle 0 \vert {\cal O} (^3P_0^{[8]})\vert  0\rangle \right ] \nonumber \\
& \qquad \qquad \qquad \times\,  \left [ C_A \int_x^1 \frac{\d  \hat x}{\hat x} \, \left  ( \frac{\hat x}{x} -1  \right )\, f_1^{g }(\hat x, \mu^2)\, + \, C_F \sum_{i = q, \bar q} \, \int_x^1 \frac{\d  \hat x}{\hat x} \, \left  ( \frac{\hat x}{x} -1  \right )\, f_1^{i }(\hat x, \mu^2)\right ] \nonumber \\
& = \sigma_{UU}^{\cos 2\phi_\psi}\, [ (\delta P_{gg}\otimes f_1^g )(x, \mu^2) \, + \,  (\delta P_{gi}\otimes f_1^i )(x, \mu^2) ] = F_{UU}^{\cos 2\phi_\psi}\,,
\end{align}
which shows the exact matching of the TMD and collinear results in the intermediate region for the structure function ${\cal F}_{UU}^{\cos2\phi_\psi}$. This is achieved without the need of any shape function because of the absence of a logarithmic term at the perturbative order we are considering. 

\section{TMD factorization and matching with the collinear framework}
\label{sec:shape}

On the basis of the above considerations,  TMD factorized expressions for the structure functions ${\cal F}_{UU,T}$ and  ${\cal F}_{UU,L}$ have to take into account smearing effects~\cite{Bacchetta:2018ivt}, encoded in the  shape function $\Delta^{[n]}$~\cite{Echevarria:2019ynx,Fleming:2019pzj}, which can be thought as a generalization of the long distance matrix elements of NRQCD in collinear factorization.   \
We start by assuming the validity of the following formulae,
\begin{align}
{\cal F}_{UU,T} & = \sum_n {\cal H}_{UU,T}^{[n]}\,{\cal C} \big [f_1^g \, \Delta^{[n]} \big ] (x, \bm q_\sT^2;\mu^2)\,,\nonumber \\
{\cal F}_{UU,L} & = \sum_n {\cal H}_{UU,L}^{[n]}\,{\cal C} \big [f_1^g \, \Delta^{[n]} \big ] (x, \bm q_\sT^2;\mu^2)\,,
\label{eq:F-conv}
\end{align}
where the ${\cal H}$ represent the hard parts, which can be calculated pertubatively. Moreover, we have introduced the transverse momentum convolution
\begin{align}
{\cal C} \big [f_1^g \, \Delta^{[n]} \big ] (x, \bm q_\sT^2; \mu^2) & = \int \d^2 \bm p_\sT  \int \d^2 \bm k_\sT \,  \delta^2 (\bm q_\sT -\bm p_\sT - \bm k_\sT  )\,  f_1^g (x, \bm p_\sT^2;\mu^2)  \, \Delta^{[n]} (\bm k_\sT^2, \mu^2)\,. 
\label{eq:conv}
\end{align}
As expected, in absence of smearing, $\Delta^{[n]} (\bm k_\sT^2;\mu^2) =  \langle 0 \vert {\cal O} (n) \, \vert 0 \rangle \, \delta^2(\bm k_\sT)$, and the  convolution in Eq.~\eqref{eq:conv} reduces to the product of the LDME $ \langle 0 \vert {\cal O} (n) \, \vert 0 \rangle$ with the gluon TMD $ f_1^g(x,  \bm q_\sT^2)$. 
Furthermore, this convolution can be expressed as follows
\begin{align}
{\cal C} \big [f_1^g \, \Delta^{[n]} \big ] (x,  \bm q_\sT^2;\mu^2) & =\int \d^2 \bm b_\sT \, e^{-i \bm b_\sT \cdot \bm q_\sT }\, \widehat f_1^g (x, \bm b_\sT^2;\mu^2)\, \widehat \Delta^{[n]} (\bm b_\sT^2,\mu^2) \,,
\label{eq:convolutions}
\end{align}
where we have introduced  the Fourier transform of the shape function,
\begin{align}
\widehat \Delta^{[n]} (\bm b_\sT^2,\mu^2)=  \frac{1}{2\pi}\int \d^2 \bm k_\sT \,e^{i \bm b_\sT\cdot \bm k_\sT} \Delta^{[n]} ( \bm k_\sT^2, \mu^2) \,.\end{align}

We are now able to show that the following expression of $ \widehat \Delta^{[n]}$, valid  at LO in $\alpha_s$,  
\begin{align}
\widehat \Delta^{[n]} (\bm b_\sT^2, \mu^2) = \frac{1}{2 \pi}\, \langle 0 \vert {\cal O}(n) \vert 0 \rangle   \left ( 1 - \frac{\alpha_s}{2 \pi} \frac{C_A}{2}\, \ln^2 \frac{\mu^2}{\mu_b^2} \right )\,,
\end{align}
which leads to 
\begin{align}
\Delta ^{[n]} ( \bm k_\sT^2,\mu^2) = \frac{\alpha_s}{2 \pi^2  \bm k_\sT^2}\, C_A \, \langle 0 \vert {\cal O}(n) \vert 0 \rangle  \,  
\ln \frac{\mu^2}{\bm k_\sT^2} \label{tailofshape}
\end{align}
in the limit $\vert \bm k_\sT  \vert \gg \Lambda_{\rm QCD}$, will solve the matching issue of the TMD and collinear results in the region $\Lambda_{\rm{QCD}} \ll \vert \bm q_\sT\vert  \ll Q$. In fact, by plugging it together with Eq.~\eqref{eq:f1-FT-final} into  Eq.~\eqref{eq:convolutions}, with the choice $\mu^2=Q^2+M_\psi^2$, we get
\begin{align}
{\cal C} \big [f_1^g \, \Delta^{[n] }\big ] (x, \bm  q_\sT^2) & =\frac{\langle 0 \vert {\cal O}(n) \vert 0 \rangle}{4 \pi^2}\int \d^2 \bm b_\sT \, e^{-i \bm b_\sT \cdot \bm q_\sT }\,  \left \{ f_1^g(x, \mu^2) - \frac{\alpha_s}{2\pi} \,\left [ \left ( \frac{ C_A}{2}\,\ln^2 \frac{\mu^2}{\mu_b^2}  -  \frac{11C_A- 2 n_f }{6}\, \ln \frac{\mu^2}{\mu_b^2}  \right )    f_1^g(x, \mu^2)  \right . \right .
\nonumber \\
& \qquad  +  \left . \left . (P_{gg} \otimes f_1^g  \,+\,  P_{gi }\otimes f_1^i) (x, \mu^2)\ln \frac{\mu^2}{\mu_b^2}\,-\,  2 \sum_a \left (  C_{g/a}^{(1)} \otimes f_1^a\right )(x, \mu^2)  \right ]  \right \} \left ( 1 - \frac{\alpha_s}{2 \pi} \frac{C_A}{2}\, \ln^2 \frac{\mu^2}{\mu_b^2} \right ) \nonumber \\
& = \frac{\langle 0 \vert {\cal O}(n) \vert 0 \rangle}{4 \pi^2}\int \d^2 \bm b_\sT \, e^{-i \bm b_\sT \cdot \bm q_\sT }\,  \left \{ f_1^g(x, \mu^2) - \frac{\alpha_s}{2\pi} \,\left [ \left ( C_A \,\ln^2 \frac{\mu^2}{\mu_b^2}  -  \frac{11C_A- 2 n_f }{6}\, \ln \frac{\mu^2}{\mu_b^2}  \right )    f_1^g(x, \mu^2)  \right . \right . \nonumber \\
& \qquad  +  \left . \left . (P_{gg} \otimes f_1^g  \,+\,  P_{gi}\otimes f_1^i) (x,\mu^2) \ln \frac{\mu^2}{\mu_b^2}\,-\, 2  \sum_a \left (  C_{g/a}^{(1)} \otimes f_1^a\right )(x, \mu_b^2)  \right ]  \right \} \nonumber \\
& =  \frac{\alpha_s}{2\pi^2  \bm q_\sT^2}\, \langle 0 \vert {\cal O}(n) \vert 0 \rangle\,\left [ \left (2 C_A  \ln \frac{\mu^2}{\bm q_\sT^2}  -  \frac{11 C_A- 2 n_f }{6}\,    \right )    f_1^g(x, \mu^2) \,+\,(P_{gg} \otimes f_1^g  \,+\,  P_{gi}\otimes f_1^i) (x,\mu^2)\right ]  \,.
\end{align}
Substituting the last formula in the LO expressions for ${\cal F}_{UU,T}$ and ${\cal F}_{UU,L}$ in Eqs.~\eqref{eq:F-conv}, we recover the correct results for $F_{UU,T}$ and $F_{UU,L}$ in Eqs.~\eqref{eq:str-funct-coll}.

Along the same lines of Ref.~\cite{Bacchetta:2019qkv}, in which a TMD factorized formula has been proposed at the twist-three level for the $\cos\phi$ asymmetry for light-hadron production in SIDIS, we can conjecture that the formulae in Eq.~\eqref{eq:F-conv} are valid to all orders in $\alpha_s$, provided one includes also the nonperturbative contributions of the TMD gluon distribution and shape function. 

At this point, it may be good to stress the difference between the shape function and the wave function of the $J/\psi$ meson. In the 
lowest-order picture a quarkonium state consists of a heavy quark and an antiquark. In the center-of-mass frame of the quarkonium the momenta $\bm{k}_1$ and $\bm{k}_2$ of the heavy quark and antiquark add up to zero. Their difference defines the relative velocity $v=|\bm{v}|$: $\bm{k} \equiv \bm{k}_1 - \bm{k}_2 = m_Q \bm{v}$, and, as already seen, NRQCD involves an expansion in $v \ll 1$. The wave function $\Phi(\bm k)$ of the quarkonium in momentum space is expected to be positronium-like, with a tail that depends on $L$: the orbital angular momentum of the quark-antiquark pair, with eigenfuction  $\Psi_{LL_z}(\bm k)$. Its Fourier transform can be conveniently written as $\widehat \Psi_{fLL_z}(\bm{r})= R_L(\vert \bm{r}\vert )\, Y_{LLz}(\theta,\phi)$, where $\bm{r} = (\vert \bm r\vert , \theta, \phi)$ in spherical coordinates, $R_L(\vert \bm r \vert )$ is the radial wave function and $Y_{LL_z} (\theta, \phi)$ is a spherical harmonic.  What enters the expressions for $L=0$ states is $\int \d^3 \bm k \, \Psi_{00}(\bm k)$, which is proportional  to $R_0(0)$.  For $L=1$ states one needs to consider the linear term $k^\alpha$ in the expansion of $\Phi(\bm k)$, hence they are related to the derivative of the radial wave function $R_1'(0)$. On the other hand, the shape function $\Delta^{[n]}$ of the $Q\overline{Q}$ system is a function of $\bm{k}_1 + \bm{k}_2$ and, at LO in the expansion in the velocity parameter $v$, is equal to a delta function in $\bm{k}_1 + \bm{k}_2$. Upon radiating additional gluons and quarks, this becomes smeared out. That is the reason why in Ref.~\cite{Bacchetta:2018ivt} the shape function was referred to as smearing function. For lack of better input, the model adopted there for this function was based on the expectations for the wave function $\Phi(\bm k)$. However, what the results presented in this paper show is that the shape function (or at least its perturbative tail) is actually independent of $L$. An $L$-independent shape or smearing function would imply that it would not be an obstacle to the extraction of the CO matrix elements from a comparison between quarkonium production and open heavy quark pair production in SIDIS as proposed in Ref.~\cite{Bacchetta:2018ivt}, implying a much more robust result. Note that the transverse momentum dependence of the shape function in Eq.\ (\ref{tailofshape}) not only implies independence from the $L$ quantum number, but actually from any quantum number of the produced quarkonium. Only the overall magnitude of the shape function is a function of these quantum numbers determined by the relevant LDME.

For the angular dependent structure function we expect the following result to hold, even if due to the absence of a collinear divergence a shape function is not strictly needed:
\begin{align}
 {\cal F}_{UU}^{\cos2\phi_\psi} & = \sum_n {\cal H}_{UU}^{\cos 2 \phi_\psi [n]}\,{\cal C} \big [ w \, h_1^{\perp g} \, \Delta_h^{[n]} \big ] (x,  \bm q_\sT^2;\mu^2)\,,
 \end{align}
where 
\begin{align}
{\cal C} \big [ w \, h_1^{\perp g} \, \Delta_h^{[n]} \big ] (x, \bm q_\sT^2;\mu^2) & = \int \d^2 \bm p_\sT  \int \d^2 \bm k_\sT \,  \delta^2 (\bm q_\sT -\bm p_\sT - \bm k_\sT  )\, w(\bm p_\sT, \bm k_\sT)\,  h_1^{\perp g} (x,  \bm p_\sT^2;\mu^2)  \, \Delta_h^{[n]} ( \bm k_\sT^2,\mu^2)\, ,
\label{eq:conv-h} 
\end{align}
with $w(\bm p_\sT, \bm k_\sT)$ being a transverse momentum dependent weight function. The shape function $\Delta_h^{[n]}$ could be in general different from $\Delta^{[n]}$: the determination of its perturbative tail would require a similar study  at higher order in $\alpha_s$. We note however that a full calculation of the cross section for  $J/\psi$ production in SIDIS at the order $\alpha^2 \alpha_s^3$, within NRQCD, is still missing~\cite{Lansberg:2019adr}.  

Based on the fact that the $p_\sT$ dependence for $h_1^{\perp\, g}$ in the gluon correlator has a rank-two tensor structure in the noncontracted transverse momentum, and unpolarized vector-meson production generally has a rank-zero structure, we consider a shape function $\Delta_h^{[n]}$ of rank zero (ignoring a possible contribution from a linearly polarized quark-pair state) and a weight function expression:
\begin{align}
 w(\bm p_\sT, \bm k_\sT) = \frac{1}{ M_p^2 \, \bm q_\sT^2}\, \left [2 (\bm p_\sT \cdot \bm q_\sT)^2 - \bm p_\sT^2\, \bm q_\sT^2 \right ] \,.
\end{align}
Furthermore, the convolution in Eq.~\eqref{eq:conv-h} can be rewritten as
\begin{align}
{\cal C} \big [ w \, h_1^{\perp g} \, \Delta_h^{[n]} \big ] (x, \bm q_\sT^2;\mu^2) &  =  \pi M_p^4 \int_0^\infty  \d \vert \bm b_\sT \vert \, \vert \bm b_\sT\vert^3\, J_2(\vert \bm b_\sT\vert \vert  \bm q_\sT\vert  )\, \widehat h_1^{\perp g\,(2)} (x, \bm b_\sT^2;\mu^2)\, \widehat \Delta_h^{[n]} (\bm b_\sT^2,\mu^2)\,,
\end{align}
where we have introduced the second derivative w.r.t.\ $\bm b_\sT^2$ of the Fourier transform of the linearly polarized gluon TMD distribution, 
\begin{align}
\widehat{h}_1^{\perp g\, (2)}(x, \bm b_\sT^2;\mu^2) &= 2 \,
\bigg(\!\!-\frac{2}{M_p^2} \frac{\partial}{\partial\bm  b_\sT^2}\bigg)^{\!2}\, \, \widehat{h}_1^{\perp g} (x, \bm b_\sT^2;\mu^2) 
=  \frac{2}{M_p^4} \int_0^{\infty} \d \vert \bm p_\sT\vert  \,
                \vert \bm p_\sT\vert\,  \frac{\bm p_\sT^2}{\bm b_\sT^2} 
J_2\big(\vert \bm b_\sT\vert  |\bm{p}_\sT|\big)\, h_1^{\perp g} \big(x, \bm p_\sT^2;\mu^2\big)\, .
\end{align}

Concerning the $L$-independence of $\Delta_h^{[n]}$ at high transverse momentum, we cannot draw any conclusion, due to the absence of a logarithmic dependence. Moreover, we cannot conclude any $L$-independence for small transverse momentum for either $\Delta^{[n]}$ or $\Delta_h^{[n]}$, but the suggestions of Ref.~\cite{Bacchetta:2018ivt} and the proposed cross-checks, do allow an experimental investigation of this. Shape functions need to be experimentally extracted, just like the LDMEs and FFs have to. The EIC can play an important role in this regard.

\section{Conclusions} 
\label{sec:conc}
Let us recapitulate the main points of this work. Our starting point is the assumption that transverse momentum dependent factorization is valid for $J/\psi$ production in SIDIS at small $q_\sT$. This Ansatz is a very reasonable one, since SIDIS for light hadrons is one of the few processes for which TMD factorization is proven at all orders in $\alpha_s$. Building further on this premise, we calculate the cross section for $e\, p \to e^\prime\, J/\psi\, X$ in two different regimes. At low $q_\sT$, the cross section is factorized in terms of TMD PDFs and generic shape functions (first introduced in Refs.~\cite{Echevarria:2019ynx,Fleming:2019pzj} as the generalization of the NRQCD LDMEs to TMD factorization), while at high $q_\sT$ the factorization involves collinear PDFs and NRQCD matrix elements. Our Ansatz of TMD factorization then requires the consistency condition that both descriptions match in the intermediate region $\Lambda_\mathrm{QCD}^2\ll \bm q_\sT^2 \ll Q^2$. The perturbative calculations for the leading-power SIDIS structure functions $F_{UU,T}$ and $F_{UU,L}$ allow, at least at LO accuracy in the strong coupling constant, to deduce from the matching the specific form of the color-octet shape function at large transverse momentum. Moreover, the angular structure function $F_{UU}^{\mathrm{cos}\,2\phi_\psi}$ matches without any dependence on a shape function whatsoever, due to the absence of a logarithmic divergence.

We therefore conclude that the assumption of TMD factorization for $J/\psi$ production in SIDIS, and the necessary (but not sufficient) condition of matching at intermediate $q_\sT$, imposes certain properties to the perturbative structure of the shape functions. In particular, by performing a study within the NRQCD framework up to the order $\alpha_s^2$, with the inclusion of CO contributions of the order $v^4$ with respect to the CS one, we find that the perturbative tails of the shape functions are independent of the quantum numbers of the intermediate CO Fock states, except for their overall magnitude given by the NRQCD LDMEs. Our conclusions of course hold for any other quarkonium state with the same quantum numbers of the $J/\psi$ meson, such as the $\psi(2 S)$ and $\Upsilon(nS)$ states. One consequence of this is that the feasibility of extracting the CO LDMEs by comparing quarkonium and open heavy-quark production at an EIC, proposed in Ref.~\cite{Bacchetta:2018ivt}, is not hampered. 

We note that our perturbative result agrees with the findings in recent work based on the soft-collinear effective theory (SCET) approach~\cite{Fleming:2019pzj}, although we are unable to draw conclusions on the nonperturbative structure of the shape functions with our method. Corroborating the results in Ref.~\cite{Fleming:2019pzj} from a different and arguably simpler approach, we thus believe that our study could contribute towards a full proof of TMD factorization for quarkonium production in SIDIS. 

\section{Acknowledgements} 
We thank Werner Vogelsang for useful discussions on the derivation of Eq.~\eqref{eq:delta-sm-qT-massless}. This project has received funding from the European Union's Horizon 2020 research and innovation programme under grant agreement No.~824093 (STRONG 2020). U.D.\ acknowledges financial support by Fondazione Sardegna under the project Quarkonium at LHC energies, CUP F71I17000160002 (University of Cagliari).
\appendix

\section{Reference frames}
\label{sec:frames}

A convenient reference frame for the calculation of the structure functions for the  process $e(\ell) + p(P) \to e(\ell^{\prime}) + J/\psi(P_\psi) + X$ is defined by adopting  light-cone coordinates with respect to the directions of the relevant hadron four-momenta, $P$ and $P_\psi$. We introduce the light-like vectors $n_+$ and $n_-$ such that $n_+\cdot n_- = 1$. Neglecting the proton mass, 
\begin{align}
n_+^\mu = P^\mu\,, \qquad  n_-^\mu = \frac{1}{P\cdot P_\psi}\,\left (P_\psi^\mu - \frac{M_\psi^2 }{2 P\cdot P_\psi}\,P^\mu \right ) \,.
\end{align}
Hence the four-momentum of the virtual photon can be written as
\begin{align}
q^\mu = - \xB\left  (1-\frac{q_\sT^2}{Q^2} \right ) n_+^\mu + \frac{Q^2}{2\xB}\, n_-^\mu + q^\mu_\sT\,,
\label{eq:qmu}
\end{align}
with  $q_\sT^2 =- \bm q_\sT^2$. From Eq.~\eqref{eq:qmu}, the off-collinearity of the process is determined as
\begin{align}
q_\sT^\mu & = q^\mu + \left ( 1 - \frac{\bm q_\sT^2}{Q^2} +\frac{M^2_\psi}{\hat z^2 Q^2} \right ) \xB P^\mu - \frac{1}{\hat z}\, P_\psi^\mu 
   = q^\mu +\left (1-\frac{\bm q_\sT^2}{Q^2} - \frac{M^2_\psi}{\hat z^2 \,Q^2} \right ) \, \hat x \, p_a^\mu  -\frac{1}{\hat z}\, P_\psi^\mu\,.
\end{align}
At the partonic level, the Mandelstam variables can be expressed as
\begin{align}
\hat s & = (q+p_a)^2 
 = Q^2 \left ( \frac{1-\hat x}{\hat x}\right )\,, \nonumber \\
\hat t & = (q-P_\psi)^2 
= -(1- \hat z) Q^2 - \hat z \bm q_\sT^2 - \frac{1-\hat z}{\hat z}\, M_\psi^2\,,  \nonumber \\
\hat u & = (p_a-P_\psi)^2
 = -\frac{\hat z}{\hat x}\, Q^2 + M_\psi^2\,.
\label{eq:mandelstam}
\end{align}
In the calculation of the cross section, we can perform the following replacement 
\begin{align}
\delta ((p_a^\prime)^2)  = \delta( (q+p_a - P_\psi)^2) & 
 = \frac{1}{\hat z Q^2}\, \delta \left (\frac{\bm q_\sT^2}{Q^2}  + \frac{1-\hat z}{\hat z^2 }\, \frac{M_\psi^2}{Q^2} -\frac{(1-\hat x)(1- \hat z) }{\hat x \hat z} \right )\,.
\end{align}
Moreover, from the kinematical constraints 
\begin{align}
\hat s \ge & M^2_\psi\,,\qquad 
 - \frac{(\hat s + Q^2) (\hat s -M_\psi^2)}{\hat s}  \le  \hat t \le 0\,, \qquad  
 -(\hat s - M^2_\psi)- Q^2 \le  \hat u \le - \frac{Q^2 M^2_\psi}{\hat s}\,, 
\end{align}
we obtain
\begin{align}
\xB &  \le \hat x   \le \frac{Q^2}{Q^2+M_\psi^2} \equiv \hat x_{\rm max}\,, \qquad  \hat z \le 1\,,
\end{align}
and, from momentum conservation, 
\begin{align}
\hat t= M_\psi^2 -Q^2 -\hat s -\hat u = - \frac{1- \hat z}{\hat x}\, Q^2\,.
\end{align}
By comparing the above relation with the second of Eqs.~\eqref{eq:mandelstam}, we get
\begin{align}
\hat t =  - \frac{1- \hat z}{\hat x}\, Q^2 = -\frac{1}{1-\hat x} \left [ \hat z \,\bm q_\sT^2  + \frac{1- \hat z}{\hat z}\, M_\psi^2 \right ] \,.
\label{eq:that}
\end{align}

Alternatively, this process can be studied in a frame where the three-momenta $\bm P$ and $\bm q$ are collinear and lie on the $z$-axis. In this frame, the virtual photon has obviously no transverse momentum. The four-momenta of the particles can be decomposed using two new vectors $\kappa_+^\mu$ and $\kappa_-^\mu$, such that $\kappa_+^2 = \kappa_-^2=0$ and $\kappa_-\cdot \kappa_+=1$,
\begin{align}
P^\mu & = \kappa^\mu_+ \,,\nonumber \\
p_a^\mu & = \frac{\xB}{\hat x}\, \kappa_+^\mu \,,\nonumber \\
q^\mu & = -\xB \kappa^\mu_+ + \frac{Q^2}{2\xB}\, \kappa_-^\mu\,,\nonumber \\
P^\mu_\psi & = \frac{M_\psi^2+\bm P_{\psi \perp}^2}{\hat z\,Q^2} \, \xB \,\kappa^\mu_+ \,+ \, \hat z\, \frac{Q^2}{2\xB}\, \kappa_-^\mu \, + \, P_{\psi \perp}^\mu\,,\nonumber \\
\ell^\mu & = \frac{1-y}{y}   \, \xB\, \kappa_+^\mu + \frac{Q^2}{2 \xB }\, \frac{1}{y}\, \kappa_-^\mu + \frac{Q}{y}\,\sqrt{1-y} \,\, \hat \ell_\perp ^\mu \,,
\label{eq:cs-pT}
\end{align}
where the following relations between the light-like vectors of the two frames hold  
\begin{align}
\kappa_+^\mu = n_+^\mu \,,\qquad \kappa_-^\mu = 2\xB^2 \frac{ \bm q_\sT^2}{Q^4}\, n_+^\mu +  n_-^\mu + \frac{2\xB}{Q^2}\, q_\sT^\mu\,.
\label{eq:light-vec}
\end{align}
The partonic Mandelstam variables in this frame read
\begin{align}
\hat s & = (q+p_a)^2 
= Q^2 \left ( \frac{1-\hat x}{\hat x}\right )\,, \nonumber \\
\hat t & = (q-P_\psi)^2 
 = - (1- \hat z) Q^2 -\frac{1- \hat z}{\hat z}\, M_\psi^2 -\frac{1}{\hat z}\, \bm P_{\psi\perp}^2 \,, \nonumber \\
\hat u & = (p_a-P_\psi)^2 
= -\frac{\hat z}{\hat x}\, Q^2 + M_\psi^2\,.
\label{eq:mandelstam2}
\end{align}
By comparing the above expression for $\hat t$ with the one in Eq.~\eqref{eq:mandelstam}, we obtain the relation between the transverse momentum of the photon $\vert \bm q_\sT\vert $ w.r.t.\ the hadrons, and the transverse momentum of the hadron 
$ \vert \bm P_{\psi \perp}\vert$ w.r.t.\ the photon and the proton,
\begin{align}
\vert \bm q_\sT \vert  = \frac{1}{\hat z}\, \vert \bm P_{\psi \perp}\vert\,.
\label{eq:qT-PhT}
\end{align}
Moreover, using the above expression together with the Sudakov decomposition of $P_\psi^\mu$ in the two frames, namely
\begin{align}
P_\psi^\mu = \frac{M_\psi^2}{\hat zQ^2}\, \xB\,n_+^\mu \, + \, \hat z\, \frac{Q^2}{2\xB}\, n_-^\mu =  \frac{M_\psi^2+\bm P_{\psi \perp}^2}{\hat z\,Q^2} \, \xB \,\kappa^\mu_+ \,+ \, \hat z\, \frac{Q^2}{2\xB}\, \kappa_-^\mu \, + \, P_{\psi \perp}^\mu\,,
\end{align}
and using Eq.~\eqref{eq:light-vec}, we obtain
\begin{align}
P_{\psi\perp}^\mu = - \hat z q_\sT^\mu - 2 \,\frac{\bm q_\sT^2}{Q^2}\, \hat z\, \xB \, n_+^\mu\,.
\end{align}

\section{Expansion of the momentum conserving delta-function for $q_\sT^2 \ll Q^2$}
\label{sec:delta-exp}

We consider the integral 
\begin{align}
I = \int_0^1\d\hat z  \,   g(\hat z) \int _0^{\hat x_{\rm max}} \d \hat x\,   f(\hat x)\, \delta (F(\hat x, \hat z))\,,
\end{align}
where $f$ and $g$ are two generic functions, 
\begin{align}
\hat x_{\rm max} = \frac{Q^2}{Q^2+M_\psi^2}\,,
\end{align}
and
\begin{align}
F(\hat x, \hat z)  & = \frac{\bm q_\sT^2}{Q^2} +  \frac{1-\hat z}{\hat z^2 }\, \frac{M_\psi^2}{Q^2} -\frac{(1-\hat x)(1- \hat z) }{\hat x \hat z} \,.
\end{align}
By introducing the variable
\begin{align}
 \hat x^\prime = \frac{\hat x}{\hat x_{\rm max}} \,, \qquad {\rm with } \qquad 0\le \hat x^\prime \le 1\,,
\end{align}
the integral becomes
\begin{align}
I = \hat x_{\rm max}\int_0^1\d\hat z  \,  \hat z^2  g(\hat z) \int _0^{1} \d \hat x^\prime\, \hat x^\prime  f(\hat x^\prime)\, \delta (G(\hat x^\prime, \hat z))\,,
\label{eq:Int}
\end{align}
with 
\begin{align}
G(\hat x^\prime, \hat z)  & =\frac{\bm q_\sT^2}{Q^2}\, \hat x^\prime \hat z^2 \,+\,   \frac{M^2_\psi}{Q^2} \, (1-\hat z) (\hat x^\prime - \hat z)   -\hat z (1-\hat z)(1-\hat x^\prime) \, .
\end{align}

After performing the first integration over $\hat x^\prime$, the integral  in Eq.~\eqref{eq:Int} can be written as
\begin{align}
I = \hat x_{\rm max} \int_{0}^{1}\d\hat z  \,  \frac{\tilde g(\hat z)}{ (1-\hat z)\left [1 + M_\psi^2/(\hat z Q^2) \,   + {\bm q_\sT^2\,  \hat z}/(Q^2 (1-\hat z))\right ]} \,  \left ( 1+ \frac{M^2_\psi}{\hat z Q^2}  \right ) \tilde f(\hat x^\prime_0)\,,
\end{align}
where
\begin{align}
\tilde g(\hat z) = \hat z \left ( 1+ \frac{M^2_\psi}{\hat z Q^2}  \right )^{-1} g(\hat z)\,,\qquad \tilde f (\hat x^\prime) =\hat x^\prime f(\hat x^\prime)\,,
\end{align}
and
\begin{align}
\hat x_0^\prime( \hat z) & =   \left ( 1+ \frac{M_\psi^2}{Q^2} \right ) \left [1 +  \frac{M_\psi^2}{\hat z Q^2} + \frac{\bm q_\sT^2}{Q^2} \, \frac{\hat z}{1-\hat z} \right ]^{-1} \,.
\end{align}

By using the identity
\begin{align}
\tilde g(\hat z) \tilde f(\hat x_0^\prime) = ( \tilde g(\hat z )- \tilde g(1)) \tilde  f(1) + \tilde g(1) \tilde f(1)   + \tilde g(\hat z ) (\tilde f(\hat x_0^\prime ) - \tilde f(1))
\end{align}
the integral in Eq.~\eqref{eq:Int} can be split in three parts, 
\begin{align}
I = \hat x_{\rm max} \,(I_1 + I_2 + I_3)\,,
\end{align}
with 
\begin{align}
I_1 & = \int_0^1\d\hat z  \,  \frac{\tilde g(\hat z) - \tilde g(1)}{ (1-\hat z)\left [1 + M_\psi^2/(\hat z Q^2) \,   +\bm q_\sT^2\, \hat z/(Q^2(1-\hat z)) \right ]} \,  \left ( 1+ \frac{M^2_\psi}{\hat z Q^2}  \right ) \tilde f(1)\,,
\end{align}
\begin{align}
I_2 & =  \tilde g(1) \int_0^{1}\d\hat z  \,  \frac{1}{ (1-\hat z)\left [1 + M_\psi^2/(\hat z Q^2) \,   +\bm q_\sT^2\, \hat z/(Q^2(1-\hat z)) \right ]} \,  \left ( 1+ \frac{M^2_\psi}{\hat z Q^2}  \right ) \tilde f(1)\,,
\end{align}
\begin{align}
I_3 & =   \int_0^{1}\d\hat z  \, \tilde g(\hat z) \, \frac{\tilde f (\hat x_0^\prime) -  \tilde f(1)}{ (1-\hat z)\left [1 + M_\psi^2/(\hat z Q^2) \,   + \bm q_\sT^2\, \hat z/(Q^2(1-\hat z))\right ]} \,  \left ( 1+ \frac{M^2_\psi}{\hat z Q^2}  \right )\,.
\end{align}

\subsection{Integral $I_1$}
In the calculation of the integral $I_1$, we can directly take the limit $q_\sT \to 0$ in the denominator of the integrand. We find 
\begin{align}
I_1 & =\int_0^{1}\d\hat z\,  \frac{\tilde g(\hat z)-\tilde g(1) }{1-\hat z} \,   \tilde f(1)\, \nonumber \\
& = \int_0^1\d\hat z\,\tilde g(\hat z)  \tilde f(1)   \frac{1 }{(1-\hat z)_+} \, \nonumber \\
& = \int_0^1\d\hat z \int_0^1 \d\hat x^\prime \,\tilde g(\hat z)  \tilde f(\hat x^\prime)  \,  \frac{1 }{(1-\hat z)_+} \delta (1-\hat x^\prime)\nonumber \\
& = \int_0^1\d\hat z \int_0^1 \d\hat x^\prime \, g(\hat z)   f(\hat x^\prime) \left ( 1+\frac{M_\psi^2}{\hat z Q^2} \right )^{-1} \frac{\hat z }{(1-\hat z)_+} \delta (1-\hat x^\prime) \nonumber \\
& = \hat x_{\rm max}^{-1}  \int_0^1\d\hat z \int_0^{\hat x_{\rm max}} \d\hat x\, g(\hat z)   f(\hat x) \left ( 1+\frac{M_\psi^2}{\hat z Q^2} \right )^{-1} \frac{\hat z }{(1-\hat z)_+} \delta (1-\hat x/\hat x_{\rm max})\,,
\end{align}
where the plus-distribution is defined in Eq.~\eqref{eq:plus}. 

\subsection{Integral $I_2$}
We can perform the integral $I_2$ exactly and then keep the leading term in the expansion in powers of $\bm q_\sT^2/Q^2$, 
\begin{align}
I_2 & =   \tilde g(1) \int_0^{1}\d\hat z  \,  \frac{1}{ (1-\hat z)\left [1 + M_\psi^2/(\hat z Q^2) \,   + \bm q_\sT^2\, \hat z/(Q^2(1-\hat z)) \right ]} \,  \left ( 1+ \frac{M^2_\psi}{\hat z Q^2}  \right ) \tilde f(1)\nonumber \\
& = \tilde g(1)\tilde f(1)\,\left [  \ln \frac{Q^2}{\bm q_\sT^2} + \ln \left ( 1+\frac{M_\psi^2}{Q^2} \right )  \right ] \nonumber \\
& \qquad  + \tilde g(1)\tilde f(1) \, \frac{\bm q_\sT^2}{Q^2} \left (1+ \frac{M_\psi^2}{Q^2} \right )^{-2} \left [  \left ( 1+ 2\,\frac{M_\psi^2}{Q^2} \right ) \ln \frac{Q^2}{\bm q_\sT^2}+  \left (  1+ 2\, \frac{M_\psi^2}{Q^2} - \frac{M_\psi^4}{Q^4} \right ) \ln \left ( 1+ \frac{M^2_\psi}{Q^2} \right ) \right . \nonumber \\
& \qquad \qquad \qquad \left .  +  \frac{M_\psi^4}{Q^4} \, \ln \frac{M_\psi^2}{Q^2} - \frac{M_\psi^2}{Q^2}\right]  \,+ \,  {\cal O}\left (\frac{\bm q_\sT^4}{Q^4}  \right ) 
\nonumber \\
& \approx  \tilde g(1)\tilde f(1)\,  \ln \bigg( \frac{Q^2+M_\psi^2}{\bm q_\sT^2}\bigg) \nonumber \\
& =  \int_0^1\d\hat z \int_0^1 \d\hat x^\prime \,\tilde g(\hat z)  \tilde f(\hat x^\prime)  \ln  \bigg( \frac{Q^2+M_\psi^2}{\bm q_\sT^2}\bigg) \delta (1-\hat x^\prime)\, \delta (1-\hat z)\nonumber \\
& =  \int_0^1\d\hat z \int_0^1 \d\hat x^\prime \, g(\hat z)   f(\hat x^\prime)  \left ( 1+\frac{M_\psi^2}{Q^2} \right )^{-1}  \ln \bigg( \frac{Q^2+M_\psi^2}{\bm q_\sT^2}\bigg) \delta (1-\hat x^\prime)\, \delta (1-\hat z)\ \nonumber \\
& =  \int_0^1\d\hat z \int_0^{\hat x_{\rm max}} \d\hat x \, g(\hat z)   f(\hat x)    \ln \bigg( \frac{Q^2+M_\psi^2}{\bm q_\sT^2}\bigg) \delta (1-\hat x/\hat x_{\rm max}) \, \delta (1-\hat z)\, .
\end{align}

\subsection{Integral $I_3$}
The integral $I_3$ is given by 
\begin{align}
I_3 & =  \int_0^{1}\d\hat z  \, \tilde g(\hat z) \, \frac{\tilde f (\hat x_0^\prime(\hat z)) -  \tilde f(1)}{(1-\hat z )\left [1 + M_\psi^2/(\hat z Q^2) \,   + \bm q_\sT^2\, \hat z/(Q^2(1-\hat z)) \right ]  } \,  \left ( 1+ \frac{M^2_\psi}{\hat z Q^2}  \right )\nonumber \\
& =  \int_0^{1}\d\hat z  \,  g(\hat z) \, \frac{\hat z}{1-\hat z}\frac{\tilde f (\hat x_0^\prime(\hat z)) -  \tilde f(1)}{\left [1 + M_\psi^2/(\hat z Q^2) \,   + \bm q_\sT^2\, \hat z/(Q^2(1-\hat z)) \right ]  } \, .
\end{align}
We can trade the integration variable $\hat z$ with $\hat x^\prime$ by inverting the relation $\hat x^\prime = x_0^\prime(\hat z)$. This amounts to solve a second order equation in $\hat z$ with solutions
\begin{align}
& \hat z_{-}  = \frac{M_\psi^2}{Q^2}\, \frac{\hat x^\prime}{M_\psi^2/Q^2 + 1-\hat x^\prime}  \left [ 1+ \frac{\bm q_\sT^2}{Q^2}\,\frac{M_\psi^2}{Q^2}\, \left (1 + \frac{M_\psi^2}{Q^2}  \right )^{-1}\, \frac{\hat x^{\prime 2}}{(M_\psi^2/Q^2 + 1-\hat x^\prime ) (1-\hat x^\prime)} \right ]+ {\cal O} \left ( \frac{\bm q_\sT^4}{Q^4}  \right ) \,,\nonumber \\
\hat z_+& = 1-  \frac{\bm q_\sT^2}{Q^2}\, \left (1+ \frac{M^2_\psi}{Q^2} \right )^{-1}\, \frac{\hat x^\prime}{ 1-\hat x^\prime } + {\cal O} \left ( \frac{\bm q_\sT^4}{Q^4}  \right ) \,.
\end{align}
The first solution is not physically acceptable, since momentum conservation implies that $q_\sT = 0$ when $\hat z = 1$ and $ 0< x^\prime < 1$.  Therefore we take $\hat z = \hat z_+$, which is the only solution surviving in the massless limit. We note that, if we neglect terms of the order $\bm q_\sT^4/Q^4$,  we obtain the  requirement $\hat x^\prime \le 1 -\bm q_\sT^2/(Q^2+M_\psi^2)$ since 
$\hat z\ge 0$. Moreover we find
\begin{align}
 \d\hat z  & = \frac{\bm q_\sT^2}{Q^2} \left (1 + \frac{M_\psi^2}{Q^2}  \right )^{-1}\,  \frac{1}{(1-\hat x^\prime )^2}\,  \d \hat x^\prime\,,\nonumber \\
&  \frac{\hat z}{(1-\hat z) \left [1 + M_\psi^2/(\hat z Q^2)\,   + \bm q_\sT^2\, \hat z/(Q^2(1-\hat z)) \right ] }   = \frac{Q^2}{\bm q_\sT^2} \,(1-\hat x^\prime)\,.
\end{align}
By substituting in the expression for $I_3$, taking into account that $g(\hat z) \to g(1)$ as $q_\sT\to 0$, we obtain
\begin{align}
I_3 & = \left ( 1+\frac{M_\psi^2}{Q^2} \right )^{-1}g(1) \int_0^1 \d \hat x^\prime \, \frac{\tilde f (\hat x^\prime) -  \tilde f(1)}{1-\hat x^\prime }\nonumber \\
& = \left ( 1+\frac{M_\psi^2}{ Q^2} \right )^{-1}g(1)    \int_0^1 \d \hat x^\prime \, \frac{\tilde f (\hat x^\prime) }{(1-\hat x^\prime)_+ }\nonumber \\
& =  \int_0^1 \d \hat z\, g(\hat z) \int_0^1 \d \hat x^\prime \, f(\hat x^\prime)\, \left ( 1+\frac{M_\psi^2}{ Q^2} \right )^{-1} \frac{\hat x^\prime}{(1-\hat x^\prime)_+}\, \delta (1-\hat z)\nonumber \\
& =  \int_0^1  \d \hat z \, g(\hat z)\,\int_0^{\hat x_{\rm max}} \d \hat x  \, f(\hat x )\,  \frac{\hat x /\hat x_{\rm max}}{(1-\hat x / \hat x_{\rm max})_+}\, \delta (1-\hat z)\,.
\end{align}

\subsection{The sum $I_1+I_2+I_3$}
Summing up the different contributions we find that the integral
\begin{align}
I = \int_0^1\d\hat z  \,   g(\hat z) \int _0^{\hat x_{\rm max}} \d \hat x\,   f(\hat x)\, \delta \left (\frac{\bm q_\sT^2}{Q^2} +  \frac{1-\hat z}{\hat z^2 }\, \frac{M_\psi^2}{Q^2} -\frac{(1-\hat x)(1- \hat z) }{\hat x \hat z} \right ) \,,
\end{align}
when $\bm q_\sT^2 \ll Q^2$, is equal to 
\begin{align}
I & = \hat x_{\rm max}  \int_0^1  \d \hat z \, g(\hat z)\,\int_0^{\hat x_{\rm max}} \d \hat x  \, f(\hat x )\, \left \{  \frac{\hat x/ \hat x_{\rm max}}{(1- \hat x/\hat x_{\rm max})_+}\, \delta (1-\hat z)  \, + \,   \frac{Q^2 + M^2_\psi}{Q^2 + M_\psi^2/\hat z}  \,  \frac{\hat z }{(1-\hat z)_+}\,  \delta (1-\hat x/\hat x_{\rm max}) \right . \nonumber \\
&  \qquad \qquad \qquad \qquad \qquad  \qquad \qquad \qquad  \qquad + \left .   \delta (1- \hat x / \hat x_{\rm max}) \,  \delta (1-\hat z) \left [ \ln \frac{Q^2}{\bm q_\sT^2} + \ln \left ( 1+\frac{M_\psi^2}{Q^2}  \right )  \right ]   \right \} \,, 
\label{eq:int-sm-qT}
\end{align}
and therefore our final result reads 
\begin{align}
\delta \left (\frac{\bm q_\sT^2}{Q^2} + \frac{1-\hat z}{\hat z^2} \, \frac{M_\psi^2}{Q^2}- \frac{(1-\hat x)(1- \hat z)}{\hat x \hat z} \right ) &  = \hat x_{\rm max} \left \{\frac{\hat x^\prime }{(1- \hat  x^\prime)_+} \, \delta (1-\hat z) \, +
\,   \frac{Q^2 + M^2_\psi}{Q^2 + M_\psi^2/\hat z}   \,  \frac{\hat z}{(1-\hat z)_+} \ \delta \left (1 -\hat x^\prime \right )  \right .\nonumber \\
&\qquad  \left .+ \,  \delta (1- \hat x^\prime) \delta (1-\hat z)  \ln \bigg(\frac{Q^2+M_\psi^2}{\bm q_\sT^2}\bigg) \right \}\,,
\label{eq:delta-sm-qT}
\end{align}
where 
\begin{align}
\hat x_{\rm max} = \frac{Q^2}{Q^2 + M_\psi^2}\,,  \qquad  \hat x^\prime = \frac{\hat x}{\hat x_{\rm max}}\,.
\label{eq:xp-def}
\end{align}
In the limit $M_\psi \to 0$,  $\hat x_{\rm max} \to 1$, $\hat x^\prime \to \hat x$ and we recover the known relation~\cite{Meng:1995yn}
\begin{align}
\delta \left (\frac{\bm q_\sT^2}{Q^2} - \frac{(1-\hat x)(1- \hat z)}{\hat x \hat z} \right ) &  = \frac{\hat x }{(1- \hat  x )_+} \, \delta (1-\hat z) \, +   \,  \frac{\hat z}{(1-\hat z)_+} \, \delta \left (1 -\hat x \right )  \,+ \,  \delta (1- \hat x) \delta (1-\hat z)  \ln \frac{Q^2}{\bm q_\sT^2} \,,
\label{eq:delta-sm-qT-massless}
\end{align}
which is valid in the region $\bm q_\sT^2\ll Q^2$. As is clear from comparing Eq.~\eqref{eq:delta-sm-qT} with \eqref{eq:delta-sm-qT-massless}, the inclusion of a heavy mass gives rise to a logarithm $\ln (Q^2+M_\psi^2)/\bm q_\sT^2$ instead of $\ln Q^2/\bm q_\sT^2$, which can be traced back to the extra term $\ln (1+ M_\psi^2/Q^2)$ in Eq. \eqref{eq:int-sm-qT}. This further supports $\mu^2=Q^2+M_\psi^2$ as the natural factorization scale for the process under study.

Finally, we note that the integration range may not start from $0$, but rather from some minimum $x$ value $a=x_{\min}$. In that case one can apply the above result to $\bar{f}(x)=f(x)\theta(x-a)$, where $\theta$ denotes the Heaviside function, for which holds (even though $\bar{f}$ is not a smooth function):
\begin{align}
\int_0^1 \mathrm{d}x \frac{\bar{f}(x)}{(1-x)_+} = \int_a^1 \mathrm{d}x \frac{f(x)-f(1)}{1-x} + f(1) \ln(1-a) \equiv \int_a^1 \mathrm{d}x \frac{f(x)}{(1-x)_+}\,,
\end{align}
see also Eq.~\eqref{eq:plus}.

\end{document}